\documentclass[12pt]{article}
\pdfoutput=1
\usepackage{graphics, color}
\usepackage{graphicx}
\usepackage{amssymb}
\usepackage{amsmath}

\usepackage{subfigure} 
\usepackage[hang,small]{caption}

\newcommand{\sect}[1]{\section{#1}\setcounter{equation}{0}}

\def\gsim{\, \rlap{$>$}{\lower 1.1ex\hbox{$\sim$}}\,}
\def\lsim{\, \rlap{$<$}{\lower 1.1ex\hbox{$\sim$}}\,}
\def\p{\partial}

 \newcommand{\be}{\begin{equation}}
\newcommand{\ee}{\end{equation}}
 \newcommand{\bal}{\begin{align}}
 \newcommand{\eal}{\end{align}}
\newcommand{\ben}{\begin{equation*}}
\newcommand{\een}{\end{equation*}}
\newcommand{\bea}{\begin{eqnarray}}
\newcommand{\eea}{\end{eqnarray}}
\newcommand{\bean}{\begin{eqnarray*}}
\newcommand{\eean}{\end{eqnarray*}}
\newcommand{\bes}{\begin{subequations}}
\newcommand{\ees}{\end{subequations}}

\usepackage{epsfig}

\newcommand{\comment}[1]{}

\begin{document}

\begin{titlepage}

\bigskip
\bigskip\bigskip\bigskip
\centerline{\Large \bf A Holographic Josephson Junction}
\bigskip\bigskip\bigskip
\bigskip\bigskip\bigskip

\centerline{{\large Gary T. Horowitz, Jorge E. Santos, and Benson Way}}
\medskip
\centerline{\em Department of Physics}
\centerline{\em University of California}
\centerline{\em Santa Barbara, CA 93106-4030}\bigskip

\bigskip
\bigskip\bigskip


\begin{abstract}
We construct a gravitational dual of a Josephson junction. Calculations on the gravity side reproduce the standard relation between the current across the junction and the phase difference of the condensate. We also study the dependence of the maximum current on the temperature and size of the junction and  reproduce familiar results. 

\end{abstract}
\end{titlepage}
\baselineskip = 17pt
\setcounter{footnote}{0}

\sect{Introduction}
One of the most remarkable results to emerge from string theory is gauge/gravity duality \cite{Maldacena98,Gubser:1998bc,Witten:1998qj} -- the equivalence between a theory of gravity and a nongravitational theory. This equivalence is called holographic since the nongravitational system lives in a lower dimensional space than the one with gravity. In recent years this duality has been successfully applied to condensed matter systems \cite{Hartnoll:2009sz,McGreevy:2009xe}. In particular, a gravitational dual of a superconductor has been found \cite{Gubser:2008px,Hartnoll:2008vx}. In this paper, we extend this construction to obtain a gravitational dual of a Josephson junction. 

A Josephson junction \cite{Josephson:1962zz} consists of two superconductors separated by a weak link.  If the link supports a phase difference $\gamma$ between the condensate of the two superconductors, there will be a current across the junction given by
\be\label{current}
J = J_{\max} \sin \gamma.
\ee
This current exists with no applied voltage.  There are several types of Josephson junctions depending on the nature of this link. It can be an insulator (SIS junctions), a normal conductor (SNS junctions), or even a very narrow superconductor (e.g. bridges and point contacts). 

We will construct a gravitational dual of an SNS Josephson junction and reproduce (\ref{current}). We will also show that $J_{\max}$ decreases exponentially with the size of the weak link as expected, and study its temperature dependence. Although our main motivation is to extend the range of condensed matter phenomena that can be described gravitationally, refinements of this approach may provide new insights into Josephson junctions made with high $T_c$ superconductors.

\sect{The Model}
We begin with the action
\be\label{action}
S=\int d^4x\;\sqrt{-g}\left[R+\frac{6}{L^2}-\frac{1}{4}F_{\mu\nu}F^{\mu\nu}-|D\psi|^2-m^2|\psi|^2\right],
\ee
where $F=dA$ and $D=\nabla-iqA$.  We will restrict our analysis to the probe approximation.  That is,  we rescale $\psi=\tilde\psi/q,$ $A=\tilde A/q$ and take $q\rightarrow\infty$, keeping $\tilde\psi$ and $\tilde A$ fixed.  In this limit, the backreaction of these fields on the metric can be ignored.  Therefore, we can fix the metric background to be the AdS planar Schwarzschild black hole:
\be
ds^2=-f(r)dt^2+\frac{dr^2}{f(r)}+r^2(dx^2+dy^2)\;,\qquad f(r)=\frac{r^2}{L^2}\left(1-\frac{r_0^3}{r^3}\right)\;,
\ee
where $r_0$ is the black hole horizon radius and $L$ the AdS length scale.
The temperature of this black hole is given by
\be
T=\frac{3r_0}{4\pi L^2}\;.
\ee

This simple model has been shown to provide a dual description of a superconductor \cite{Hartnoll:2008vx}.  As a consequence of a scaling symmetry, the critical temperature $T_c$ of the superconductor is proportional to the chemical potential $\mu$.  We will allow the chemical potential to have spatial dependence\footnote{For other inhomogeneous holographic superconductors, see \cite{Keranen:2009ss,Flauger:2010tv}, and see \cite{Aperis:2010cd} for a related solution.} $\mu(\vec x)$ so that a fixed temperature $T$ is above the critical temperature in a narrow gap but below the critical temperature elsewhere. In this way, we model two superconductors separated by a normal phase.  

Without loss of generality, we will create our junction along the $x$ direction.  Our fields will now be functions of $r$ and $x$.  A Josephson junction requires a phase difference, so we must include a phase in our scalar field.  Therefore, we consider solutions of the form
\be
\tilde\psi=|\psi|e^{i\varphi}\;,\qquad \tilde A=A_t\;dt+A_r\;dr+A_x\;dx\;,
\ee
where $|\psi|$, $\varphi$, $A_t$, $A_r$, and $A_x$ are all real functions of $r$ and $x$.   Instead of $A$, we wish to work with the gauge-invariant fields $M=A- d\varphi$.  

The equations of motion are
\begin{subequations}
\begin{align}
\p_{r}^2|\psi|+\frac{1}{r^2f}\p_{x}^2|\psi|+\left(\frac{f'}{f} + \frac{2}{r}\right)\p_r|\psi|+\frac{1}{f}\left(\frac{M_t^2}{f}-fM_r^2-\frac{M_x^2}{r^2}-m^2\right)|\psi|&=0\;,\label{psi}\\
\p_{r}^2M_t+\frac{1}{r^2f}\p_{x}^2M_t+\frac{2}{r}\p_r M_t-\frac{2|\psi|^2}{f}M_t&=0\;,\label{Mt}\\
\p_{x}^2M_r-\p_{r}\p_xM_x-2r^2|\psi|^2M_r&=0\;,\label{Mr}\\
\p_{r}^2M_x-\p_{r}\p_xM_r+\frac{f'}{f}\left(\p_r M_x-\p_x M_r\right)-\frac{2|\psi|^2}{f}M_x&=0\;,\label{Mx}\\
\p_{r}M_r+\frac{1}{r^2f}\p_xM_x+\frac{2}{|\psi|}\left(M_r\p_r|\psi|+\frac{M_x}{r^2f}\p_x|\psi|\right)+\left(\frac{f'}{f} + \frac{2}{r}\right)M_r&=0\;.\label{constraint}
\end{align}
\label{tot_equations}
\end{subequations}
Note that the phase $\varphi$ no longer appears in the equations of motion. The first four equations (\ref{psi}-\ref{Mx}) are second-order dynamical equations while the last one (\ref{constraint}) is a first-order constraint equation coming from the conservation of the source in Maxwell's equations.  

Now we discuss the boundary conditions, beginning with $r=\infty$.  At this point, we will work in units in which $L=1$.  For simplicity, we choose\footnote{Our analysis can be extended to any mass above the Breitenl\"ohner-Freedman bound, $m^2>-9/4$.} $m^2=-2$.   With this choice of mass, the scalar has the following asymptotic form near $r=\infty$:
\be
|\psi|=\frac{\psi^{(1)}(x)}{r}+\frac{\psi^{(2)}(x)}{r^2}+O\left(\frac{1}{r^3}\right)\;.
\ee
The $\psi^{(1)}$ and $\psi^{(2)}$ terms are both normalizable, so we have a choice of boundary conditions.  Here, we only consider the case $\psi^{(1)}=0.$  Via gauge/gravity duality,  $\psi^{(2)}$ then gives the expectation value of a dimension two operator in the boundary field theory
\be
\langle\mathcal{O}\rangle=\psi^{(2)}.
\ee
We interpret this operator as the superconducting condensate.  

The Maxwell fields take the asymptotic form
\begin{align}
M_t&=\mu(x)-\frac{\rho(x)}{r}+O\left(\frac{1}{r^2}\right)\;,\\
M_r&=O\left(\frac{1}{r^3}\right)\;,\\
M_x&=\nu(x)+\frac{J}{r}+O\left(\frac{1}{r^2}\right)\; \label{defnu}.
\end{align}
The quantities $\mu(x)$, $\rho(x)$, $\nu(x)$ and $J$ are interpreted in the boundary field theory as the chemical potential, charge density, superfluid velocity, and current, respectively.  By solving the equations of motion near $r=\infty$, it can be shown that $J$ must be a constant. This condition is imposed by the constraint (\ref{constraint}), which in turn enforces current conservation. Our boundary conditions are determined by choosing $J$ and $\mu(x)$.  

At the horizon $r=r_0$, regularity requires $M_t=0$.  This and the equations of motion place boundary conditions on the remaining functions.  

At $x=\pm\infty$, we require that the functions approach the homogeneous ($x$-independent) solution.  We solve the ordinary differential equations numerically, imposing regularity at the horizon and choosing $J$ and $\mu(\infty) = \mu(-\infty)$ at $r=\infty$.  With non-zero\footnote{For a discussion of holographic superconductors with nonzero current, see \cite{Basu:2008st,Arean:2010xd,Sonner:2010yx}} $J$, there are typically two solutions.  We choose the solution with the lower free energy, which is also the solution that has a larger value of the condensate \cite{Arean:2010xd}.  

  The gauge invariant phase difference is $\gamma = \Delta \varphi -\int A_x$ where the integral is across the gap. Since the edges of our gap will not be completely sharp (for numerical reasons), it is more convenient to use (\ref{defnu}) and set
\be
\gamma=- \int_{-\infty}^\infty dx \;[\nu(x) -\nu(\pm\infty)]\;.
\label{gamma}
\ee
The second term is needed as a regulator since the homogeneous superconductor at $x = \pm \infty$ has a nonzero current and hence a nonzero superfluid velocity.  Note that the value of this term at $+\infty$ and $-\infty$ is the same. For numerical convenience, and because the phase difference $\gamma$ between the two superconductors is a non-local quantity, we will reproduce (\ref{current}) by first picking $J$ and then computing $\gamma$ using (\ref{gamma}).

The equations of motion (\ref{tot_equations}) are invariant under the following scaling symmetry:
\be\label{scaling}
r\rightarrow ar\;,\quad (t,x,y)\rightarrow(t,x,y)/a\;,\quad M_r\rightarrow M_r/a\;,\quad(M_t,M_x)\rightarrow a(M_t,M_x)\;.
\ee
For homogeneous superconductors, the scale invariant temperature is $T/\mu$, so changing $\mu$ is equivalent to changing $T$. In the inhomogeneous case, $T$ must still be constant, even though $\mu = \mu(x)$. The parameter $a$ in (\ref{scaling}) must be constant, so there is no symmetry which exchanges an inhomogeneous chemical potential with an inhomogeneous temperature.  The scale invariant temperature is $T/T_c$ where $T_c$ is the critical temperature of the junction, which  is identical to the critical temperature of a homogenous superconductor with zero current and $\mu = \mu(\infty) = \mu(-\infty)$. $T_c$ is proportional to $\mu(\infty)$ and given by
\be\label{T_c}
T_c\approx0.0588\,\mu(\infty)\;.
\ee
When performing numerics, we will use  (\ref{scaling}) to set $r_0=1$, and adjust $T/T_c$ by varying $\mu(\infty)$.  
 
We would like to choose a profile for $\mu(x)$ such that our system resembles an SNS Josephson junction.   Therefore, we choose $\mu(x)$ to be small and approximately constant for $x\in(-\tfrac{\ell}{2},\tfrac{\ell}{2})$ and then rise rapidly to $\mu(\infty)$ outside this gap.  Inside the gap, the effective critical temperature is
\be\label{T_0}
T_0 \approx0.0588\,\mu(0)\;,
\ee
For temperatures $T_0 < T< T_c$, the gap is a  normal conductor and the region outside the gap is superconducting, so we indeed have an SNS Josephson junction.   For temperatures smaller than $T_0$, our material is everywhere in the superconducting phase, and we have an S-S'-S junction.  For $T>T_c$, our material is entirely in the normal phase.

  A profile that fits this description is given by
\be\label{profile}
\mu(x)=\mu_\infty\left\{1-\frac{1-\epsilon}{2\tanh(\frac{\ell}{2\sigma})}\left[\tanh\left(\frac{x+\tfrac{\ell}{2}}{\sigma}\right)-\tanh\left(\frac{x-\tfrac{\ell}{2}}{\sigma}\right)\right]\right\}\;,
\ee
where $\mu_\infty$ is the chemical potential at $x=\pm\infty$, and $\ell$ is the width of our junction\footnote{We have tried a different profile which is less steep and consequently easier for numerics.  This profile was also able to reproduce (\ref{current}).}.  The quantities $\sigma$ and $\epsilon$ control the steepness and depth of our profile, respectively.  For $\ell \gg \sigma$, $\mu(x)$ is quite flat inside the junction (see Fig. 1). For this profile, $T_0=\epsilon\,T_c$.

\sect{Numerics and Results}
We can now attempt to solve the coupled system of nonlinear partial differential equations (\ref{tot_equations}) numerically for any value of $J$ and any parameters in (\ref{profile}). In order to proceed, it is convenient to impose boundary conditions at $x=0$ and $x=+\infty$ rather than at $x=\pm\infty$.  By using the equations of motion and their perturbations, one can prove that our boundary conditions require that $M_r$ be odd and $|\psi|$, $M_t$, and $M_x$ be even.  This in turn imposes, at $x=0$, Dirichlet boundary conditions on $M_r$ and Neumann boundary conditions on the other functions.  We can also compactify both the radial coordinate $r$ and boundary coordinate $x$ using the change of variables $z = 1-r_0/r$ and $\tilde{x} = \tanh[x/(4 \sigma)]$. 

Our numerical method relies on a standard relaxation procedure, combined with spectral methods defined on a Chebyshev grid. Due to the exponential convergence characteristic of spectral methods on such a grid, we only require a few points in our spatial discretization scheme. In all the plots in this manuscript, we use $41$ points along the $\tilde{x}$ direction and $25$ along $z$. We have varied the number of points, and found good agreement with the aforementioned exponential convergence.

\begin{figure}[t]
\centerline{\includegraphics[width=.45\textwidth]{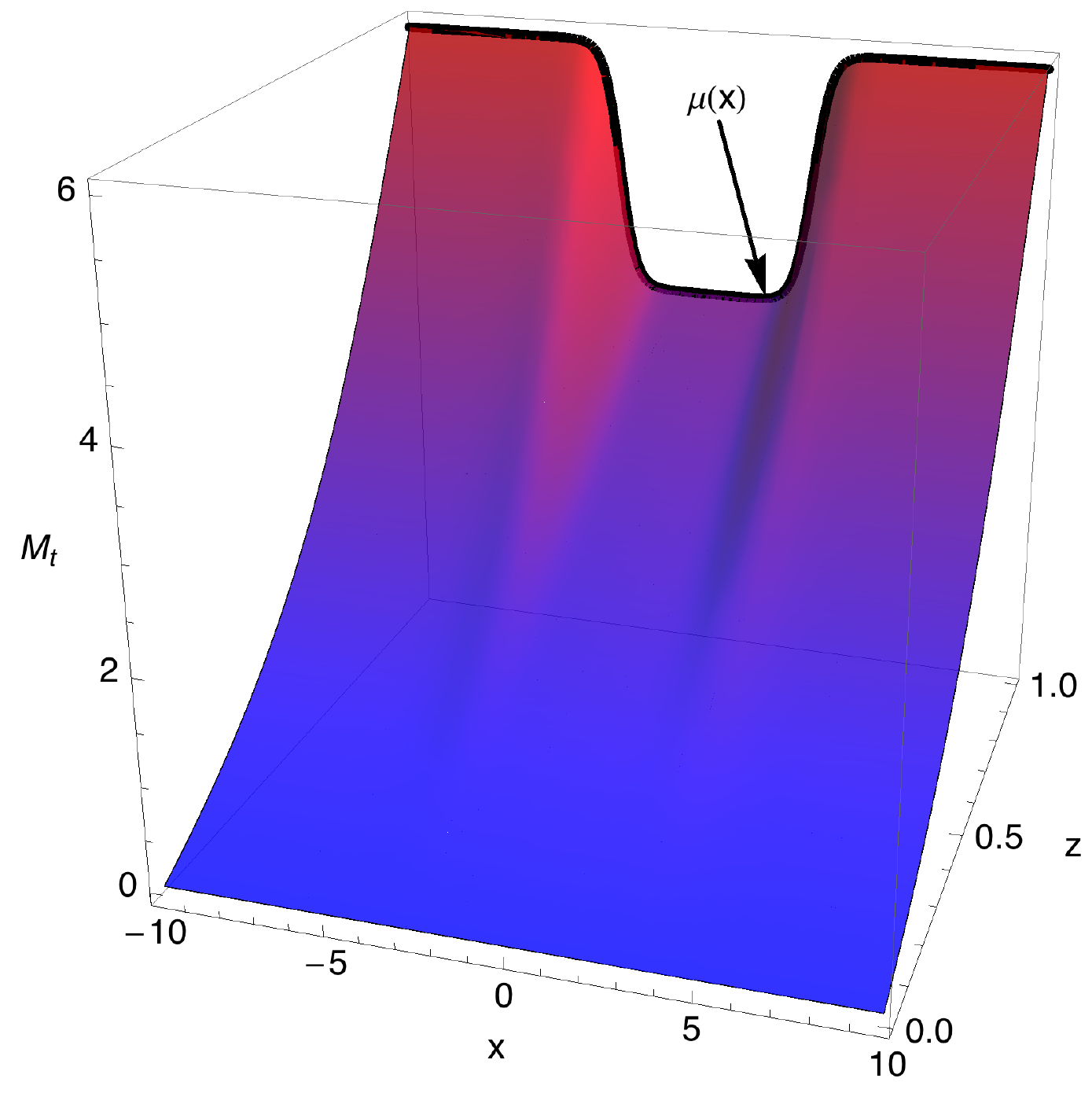}
\hspace{1cm}\includegraphics[width=.45\textwidth]{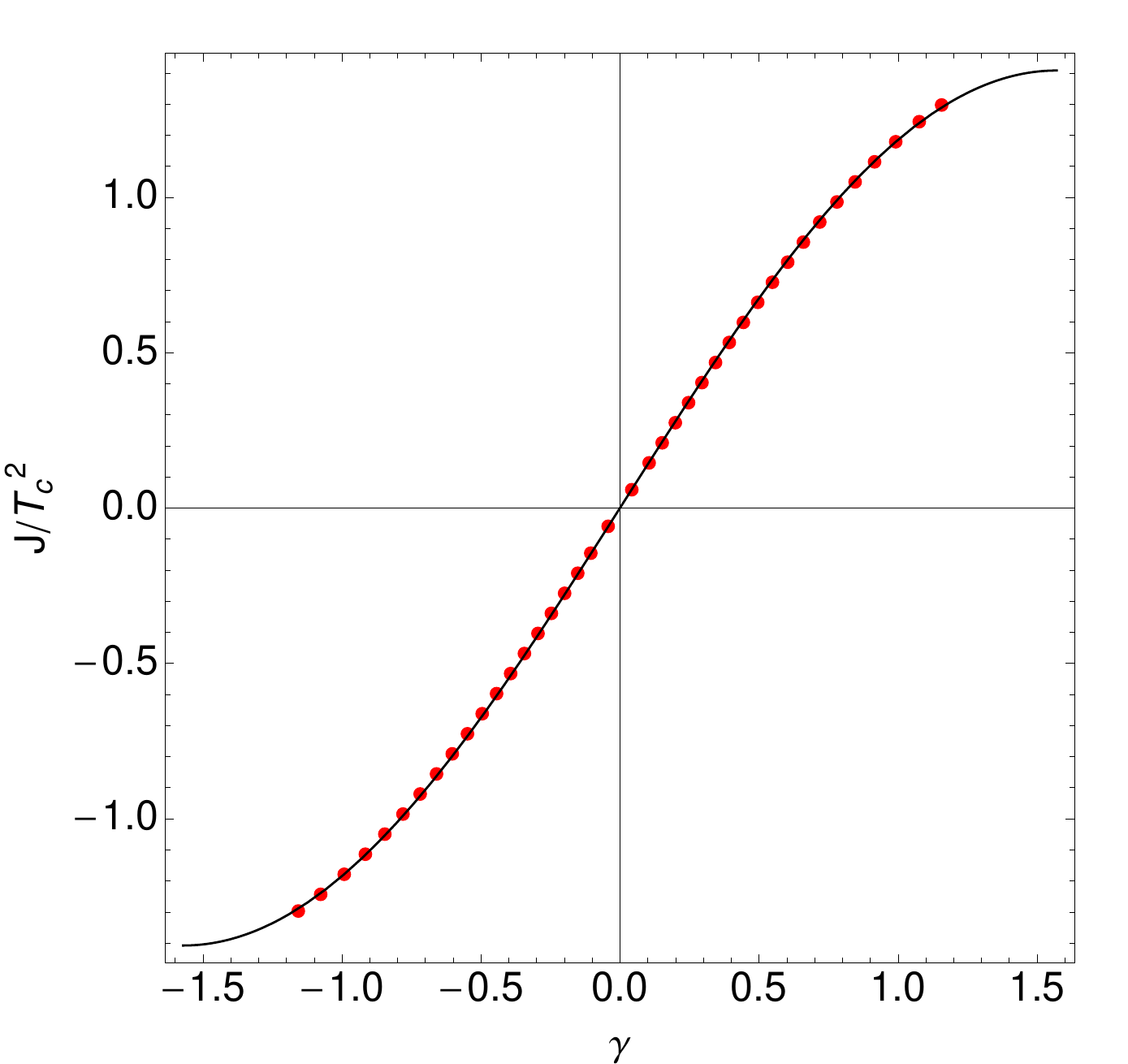}}
\caption{In the first graph, we represent $M_t$ as a function of $x$ and $z$ for $J/T^2_c \approx 0.0081$, showing the boundary profile $\mu(x)$ imposed at $z=1$. The second graph shows how $J/T^2_c$ varies as a function of $\gamma$. The solid line is the best fit sine curve.  In both plots we use $\mu_\infty = 6$, $\epsilon = 0.6$,  and $\sigma = 0.5$. The first plot has $\ell = 6 $ and the second has $\ell = 3 $.}
\label{fig:1}
\end{figure}

After solving the equations, the phase difference can be obtained through (\ref{gamma}).  We can study the dependence on the size of our junction and on the temperature by varying $\ell$ and $\mu_\infty$, respectively. As an example, we show on the left panel of Fig.~\ref{fig:1} a typical result of our numerical code for $M_t$, showing the profile (\ref{profile}) imposed on the boundary.

A standard property of Josephson junctions is (\ref{current}). We only consider phase differences that lie in the interval $(-\pi/2,\pi/2)$ since, within a cycle, the other phase differences correspond to dynamically unstable junctions \cite{barone:1982bb}. By feeding into our code successive values of $J$ and computing $\gamma$ using (\ref{gamma}), we constructed the graph on the right panel of Fig.~\ref{fig:1}. The black solid line represents the best fit of our numerical data to (\ref{current}). The agreement is striking, and predicts a maximum current across the junction of $J_{\max}/T^2_c \approx 1.408$. We have used other profiles for $\mu(x)$ to model the junction, and the resulting curve shares the same features as the right panel of Fig.~\ref{fig:1}, showing that (\ref{current}) is a robust signature of holographic SNS Josephson junctions. Two natural questions, which we address next, is how $J_{\max}$ varies with the junction width $\ell$ and the temperature $T/T_c$.

\begin{figure}[t]
\centerline{\includegraphics[width=.45\textwidth]{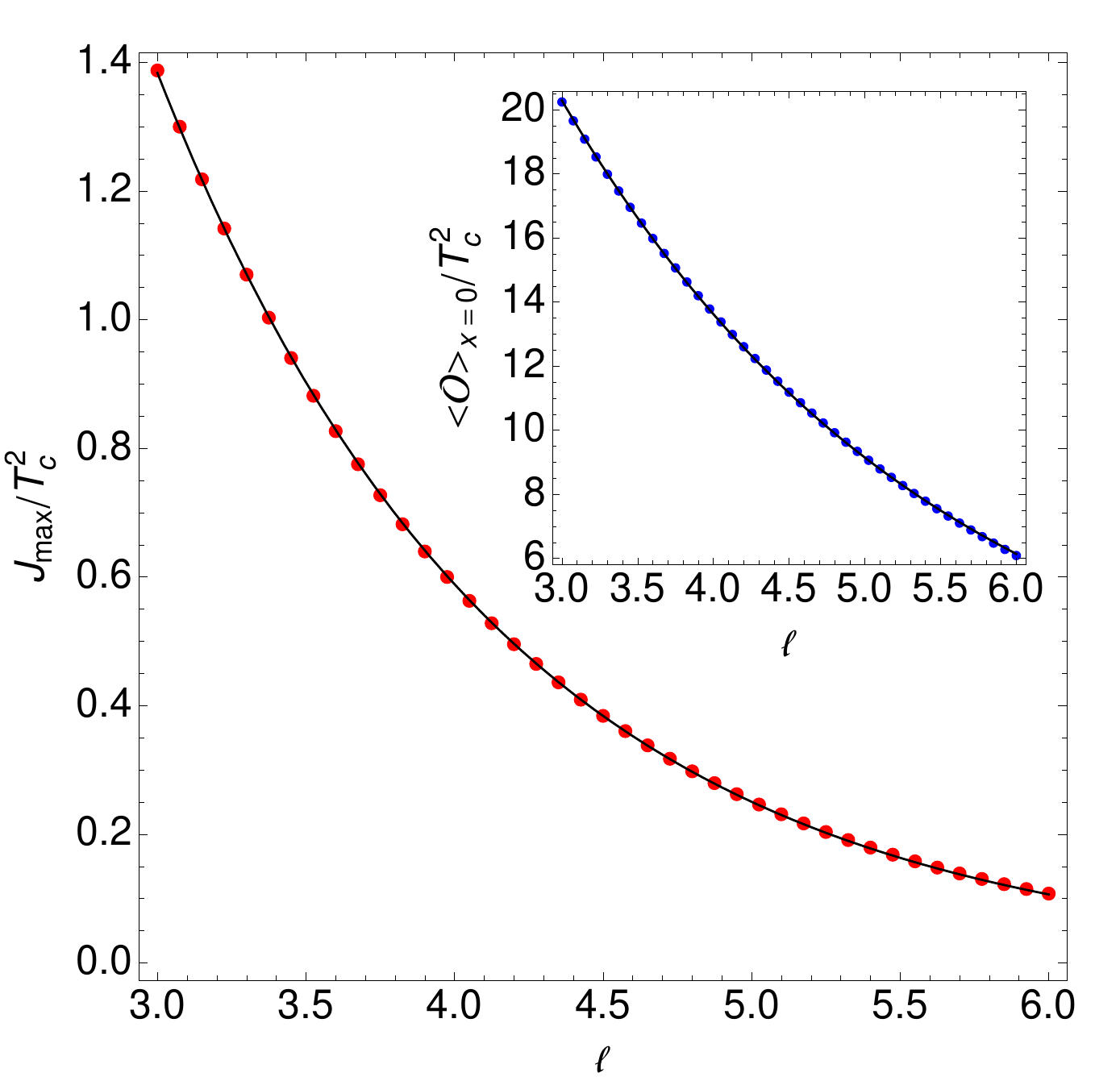}
\hspace{1cm}\includegraphics[width=.45\textwidth]{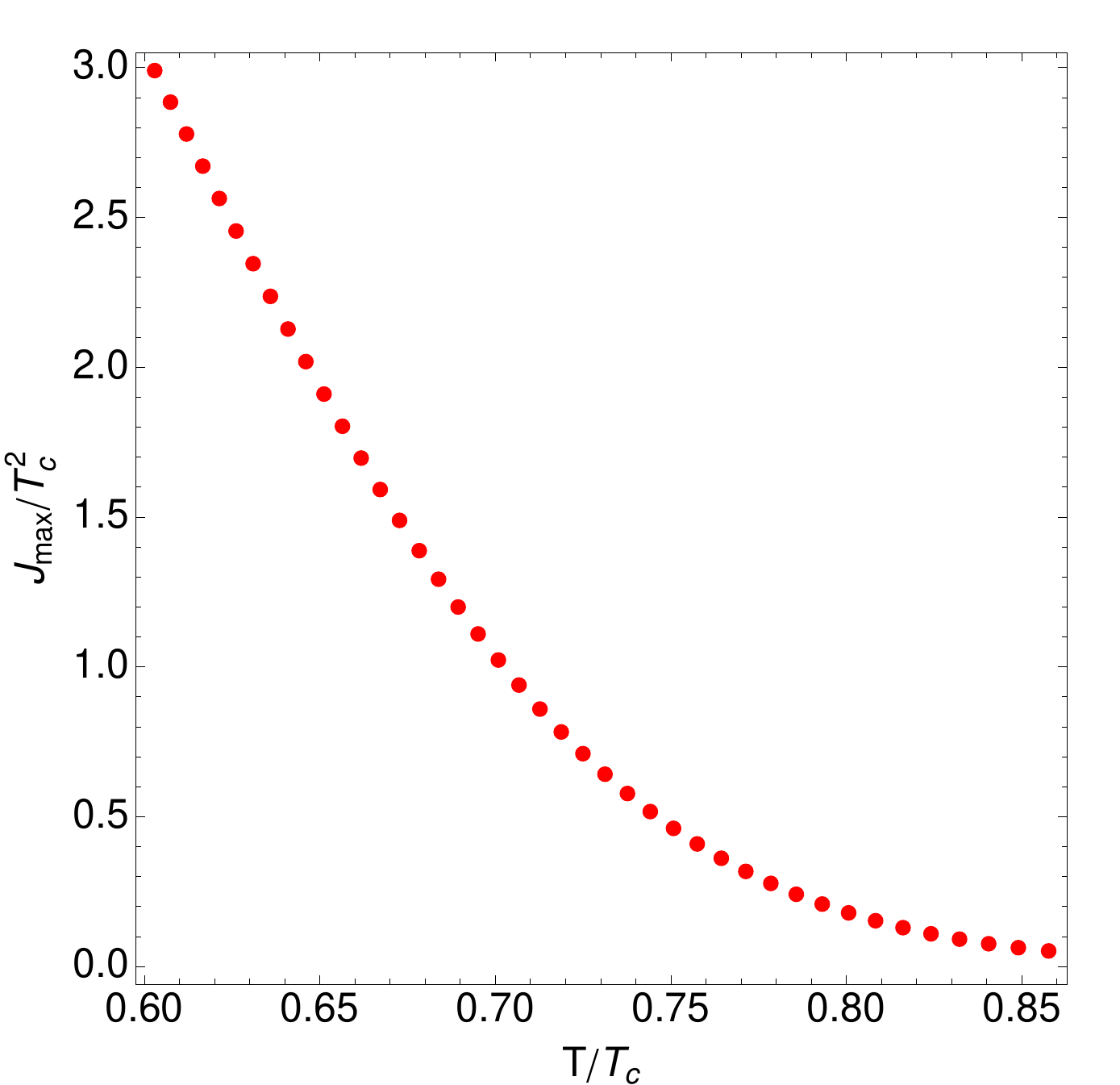}}
\caption{In the first graph, we represent $J_{\max}$  and  $\langle\mathcal{O}\rangle_{x=0}$ (inset plot) as functions of $\ell$ for $\mu_\infty = 6.0$. The second graph shows how $J_{\max}/T^2_c$ varies as a function of $T/T_c$, for $\ell = 3$. In all the plots, we use $\epsilon = 0.6$ and $\sigma = 0.5$. The black solid lines are the best fit exponential curves.}
\label{fig:2}
\end{figure}

The dependence of $J_{\max}$ on $\ell$ is shown on the left panel of Fig.~\ref{fig:2}. Once again, this curve is in good agreement with condensed matter physics \cite{tinkham:1975bb}, which for SNS junctions, predicts an exponential decay with growing $\ell$ in the maximum current:
\be
J_{\max}/T^2_c = A_0\,e^{-\frac{\ell}{\xi}}.\label{jmaxell}
\ee
This universal law is only expected to hold as long as $\xi \ll \ell$. Furthermore, a similar behavior is expected of the condensate within the barrier at zero current:
\be
\qquad \langle \mathcal{O}\rangle_{x = 0,J=0}/T^2_c= A_1\,e^{-\frac{\ell}{2\,\xi}}.\label{oell}
\ee
Note that the value of the coherence length $\xi$ should be the same in (\ref{jmaxell}) and (\ref{oell}).  We have developed an independent code that studies the limiting case of zero current, in which $|\psi|$ and $M_t$ are the only non-zero variables. We show, as an inset plot of the left panel of Fig.~\ref{fig:2}, how $\langle \mathcal{O}\rangle_{x = 0,J=0}$ varies with $\ell$. We fit both sets of data to (\ref{jmaxell}) and (\ref{oell}) and find $\{\xi ,A_0 \}  \approx \{1.17,17.96\}$  and $\{\xi,A_1\} \approx \{1.26,33.52\}$ for $J_{\max}/T_c^2$ and $\langle \mathcal{O}\rangle_{x = 0,J=0}/T^2_c$, respectively. Note that these two sets of numbers were computed using different numerical codes, each depending on a different number of variables, and each having their own numerical error.  Furthermore, the normal and superconducting phases in our junction are not cleanly separated -- our numerics prevent us from choosing a profile for $\mu(x)$ that is too steep.  We believe this justifies the $7\%$ disagreement between the two estimates of $\xi$.

We now turn our attention to the dependence of $J_{\max}$ on the temperature.  Since there is not a simple form for the expected behavior of $J_{\max}(T)$, we do not include fits to our plot on the right side of Fig.~\ref{fig:2}.

There are two distinct regimes for how $J_{\max}$ varies with $T$, corresponding to the two temperature scales (\ref{T_c}) and (\ref{T_0}) set by the profile (\ref{profile}). For temperatures smaller than $T_0$, our material is everywhere in the superconducting phase and we have an S-S'-S junction.  In this regime, we find that our system deviates far from (\ref{current}).  Indeed, this behavior is expected of S-S'-S junctions \cite{Likharev:1979zz}, but the deviation creates difficulties in finding $J_{\max}$ so we do not include that region in our plot.  In Fig.~\ref{fig:2}, we have chosen $\epsilon=0.6$, so this region corresponds to $T/T_c<0.6$.  

There is a complementary temperature regime near $T_c$, corresponding to the critical temperature at which the entire junction is in the normal phase.  Although this is an interesting regime, it is difficult to probe numerically due to large scale separations between the condensate size $\langle\mathcal O\rangle$ and the charge density $\rho$.  However, we observe that as $T$ approaches $T_c$,  $J_{\max}$ approaches zero, as is expected. 

\sect{Discussion}

We have seen that the simplest model of a holographic superconductor can also be used to construct a holographic Josephson junction by choosing $\mu(x)$ appropriately. Since the chemical potential is not constant, one might wonder why we find stationary solutions. After all, there is a nonzero electric field $E(x) = \mu'(x)$ on the boundary near the edges of the gap which might be expected to make $\rho$ time dependent. However this  effect can be balanced by diffusion \cite{Amado:2009ts} leading to the stationary configuration we find.\footnote{We thank T. Faulkner for suggesting this.} We expect that holographic Josephson junctions  with constant $\mu$ can be constructed by modifying the boundary condition for the scalar field in a narrow gap along the lines discussed in \cite{Faulkner:2010gj}.

We have restricted attention to the DC Josephson effect. The AC effect says that if a voltage difference $V$ is applied across the junction, then the solution is indeed time dependent and the phase difference satisfies
\be\label{AC}
d\gamma/dt = qV
\ee
so one gets an alternating current.  The added time dependence makes it harder to construct the appropriate gravitational solutions. However, there is a good reason to expect that the AC Josephson effect will also be reproduced on the gravity side: it can be obtained by doing a bulk gauge transformation on half of the solution. Adding a voltage difference V corresponds to adding a constant $V$ to $A_t$, but this is gauge equivalent to changing the phase of our charged scalar $\psi$ by  $qVt$. This is precisely (\ref{AC}). 

Another natural extension of our model is to add a magnetic field. This should cause the phase to increase linearly in a spatial direction along the junction. The dual gravitational solutions now depend on three variables. We leave a discussion of their properties for future work. 

Our analysis has been restricted to the probe approximation.  To go beyond the probe limit and study the effects of backreaction, one must solve the full dynamical equations, including Einstein's equations. The first step in this direction was done in  \cite{Sonner:2010yx}, where the homogeneous superconductor with nonzero current was constructed.  The backreacted junction would require solving a set of nine coupled, nonlinear partial differential equations. 

Our model can be extended to accommodate a junction whose weak link is a narrow superconductor (a Dayem bridge).  One would need a profile $\mu(x,y)$ that breaks superconductivity in the regions $x\in(-\tfrac{\ell}{2},\tfrac{\ell}{2})$, $|y|>\tfrac{h}{2}$.  As in the case with a magnetic field, the gravitational solutions will now depend on three variables.  
We expect these junctions to support a larger $J_{\max}$.  

\vskip .5cm
\centerline{\bf Acknowledgements}

It is a pleasure to thank T. Faulkner, A. Ludwig, D. Marolf, and D. Scalapino for discussions. This work was supported in part by the  National Science Foundation under Grant No.~PHY08-55415.


\begin{thebibliography}{99}
\bibitem{Maldacena98}
J. M. Maldacena, ``The large N limit of superconformal field theories and supergravity,"
Adv. Theor. Math. Phys. \textbf{2}, 231 (1998)  [Int. J. Theor. Phys. \textbf{38},
1113 (1999)] [arXiv:hep-th/9711200].
	
\bibitem{Gubser:1998bc}
  S.~S.~Gubser, I.~R.~Klebanov and A.~M.~Polyakov,
  ``Gauge theory correlators from non-critical string theory,''
  Phys.\ Lett.\  B {\bf 428}, 105 (1998)
  [arXiv:hep-th/9802109].
  
\bibitem{Witten:1998qj}
  E.~Witten,
  ``Anti-de Sitter space and holography,''
  Adv.\ Theor.\ Math.\ Phys.\  {\bf 2}, 253 (1998)
  [arXiv:hep-th/9802150].
  
\bibitem{Hartnoll:2009sz}
  S.~A.~Hartnoll,
  ``Lectures on holographic methods for condensed matter physics,''
  Class.\ Quant.\ Grav.\  {\bf 26}, 224002 (2009).
  [arXiv:0903.3246 [hep-th]].
  
  \bibitem{McGreevy:2009xe}
  J.~McGreevy,
  ``Holographic duality with a view toward many-body physics,''
  arXiv:0909.0518 [hep-th].

\bibitem{Gubser:2008px}
  S.~S.~Gubser,
  ``Breaking an Abelian gauge symmetry near a black hole horizon,''
  Phys.\ Rev.\  {\bf D78}, 065034 (2008).
  [arXiv:0801.2977 [hep-th]].
\bibitem{Hartnoll:2008vx}
  S.~A.~Hartnoll, C.~P.~Herzog, G.~T.~Horowitz,
  ``Building a Holographic Superconductor,''
  Phys.\ Rev.\ Lett.\  {\bf 101}, 031601 (2008)
  [arXiv:0803.3295 [hep-th]];
	 ``Holographic Superconductors,Ó 
	 JHEP {\bf 0812}, 015 (2008)
  [arXiv:0810.1563 [hep-th]].	

\bibitem{Josephson:1962zz}
  B.~D.~Josephson,
  ``Possible new effects in superconductive tunnelling,''
  Phys.\ Lett.\  {\bf 1}, 251 (1962).
   
\bibitem{Keranen:2009ss}
  V.~Keranen, E.~Keski-Vakkuri, S.~Nowling {\it et al.},
  ``Inhomogeneous Structures in Holographic Superfluids: I. Dark Solitons,''
  Phys.\ Rev.\  {\bf D81}, 126011 (2010)
  [arXiv:0911.1866 [hep-th]];
  ``Solitons as Probes of the Structure of Holographic Superfluids,''
  [arXiv:1012.0190 [hep-th]].
  
\bibitem{Flauger:2010tv}
  R.~Flauger, E.~Pajer, S.~Papanikolaou,
  ``A Striped Holographic Superconductor,''
   [arXiv:1010.1775 [hep-th]].
   
\bibitem{Aperis:2010cd}
  A.~Aperis, P.~Kotetes, E.~Papantonopoulos {\it et al.},
  ``Holographic Charge Density Waves,''
    [arXiv:1009.6179 [hep-th]].
   
\bibitem{Basu:2008st}
  P.~Basu, A.~Mukherjee, H.~-H.~Shieh,
  ``Supercurrent: Vector Hair for an AdS Black Hole,''
  Phys.\ Rev.\  {\bf D79}, 045010 (2009);
  [arXiv:0809.4494 [hep-th]].
  C.~P.~Herzog, P.~K.~Kovtun, D.~T.~Son,
  ``Holographic model of superfluidity,''
  Phys.\ Rev.\  {\bf D79}, 066002 (2009).
  [arXiv:0809.4870 [hep-th]].
   
\bibitem{Arean:2010xd}
  D.~Arean, M.~Bertolini, J.~Evslin and T.~Prochazka,
  ``On Holographic Superconductors with DC Current,''
  JHEP {\bf 1007}, 060 (2010)
  [arXiv:1003.5661 [hep-th]].

\bibitem{Sonner:2010yx}
  J.~Sonner, B.~Withers,
  ``A gravity derivation of the Tisza-Landau Model in AdS/CFT,''
  Phys.\ Rev.\  {\bf D82}, 026001 (2010).
  [arXiv:1004.2707 [hep-th]].

\bibitem{tinkham:1975bb}
M. Tinkham, {\em Introduction to Superconductivity},
McGraw Hill,  (1975).

\bibitem{barone:1982bb}
A. Barone, G. Patern\`o, {\em Physics and Applications of the Josephson Effect}, 
John Wiley \& Sons, New York (1982).

\bibitem{Likharev:1979zz}
  K.~K.~Likharev,
  ``Superconducting weak links,''
  Rev.\ Mod.\ Phys.\  {\bf 51}, 101 (1979).
  
\bibitem{Amado:2009ts}
  I.~Amado, M.~Kaminski, K.~Landsteiner,
  ``Hydrodynamics of Holographic Superconductors,''
  JHEP {\bf 0905}, 021 (2009).
  [arXiv:0903.2209 [hep-th]].
  
\bibitem{Faulkner:2010gj}
  T.~Faulkner, G.~T.~Horowitz, M.~M.~Roberts,
  ``Holographic quantum criticality from multi-trace deformations,''
  [arXiv:1008.1581 [hep-th]].
  

\end{thebibliography}
\end{document}